\documentclass{article}
\usepackage{spconf,amsmath,graphicx}

\usepackage{enumitem}
\setlist{nosep, leftmargin=14pt}

\usepackage{mwe} 
\usepackage{tikz}
\usetikzlibrary{positioning, shapes, arrows.meta}
\usetikzlibrary{calc}
\usetikzlibrary{positioning, shapes, arrows.meta, calc, fit}
\usetikzlibrary{backgrounds}
\usepackage{bm} 
\usepackage{bbold}
\usepackage{algorithm}
\usepackage{algpseudocode}
\usepackage{booktabs}
\usepackage{siunitx}
\usepackage{float}

\title{End-to-end optimization of sparse ultrasound linear probes}
\name{Sergio Urrea $^{\dagger}$ \qquad Adrian Basarab$^{\star}$ \qquad Hervé Liebgott$^{\star}$ \qquad Henry Arguello$^{\ddagger}$}

\address{$^{\dagger}$ Department of Electrical, Electronics, and Telecommunications Engineering, \\$\ddagger$ Department of Systems Engineering and Informatics,\\ Universidad Industrial de Santander. \\
    $^{\star}$ CREATIS, Universite Claude Bernard Lyon 1,\\ INSA-Lyon, CNRS, INSERM, UMR5220, U1294, France}
\begin{document}
%
\maketitle
\begingroup
\renewcommand\thefootnote{}
\footnotetext{%
\textcopyright\ 2026 IEEE. Personal use of this material is permitted. Permission from IEEE must be obtained for all other uses, in any current or future media, including reprinting/republishing this material for advertising or promotional purposes, creating new collective works, for resale or redistribution to servers or lists, or reuse of any copyrighted component of this work in other works.

Accepted for publication in the IEEE International Symposium on Biomedical Imaging (ISBI 2026).}
\addtocounter{footnote}{-1}
\endgroup
\begin{abstract}
Ultrasound imaging faces a trade-off between image quality and hardware complexity caused by dense transducers. Sparse arrays are one popular solution to mitigate this challenge.  
This work proposes an end-to-end optimization framework that jointly learns sparse array configuration and image reconstruction.  
The framework integrates a differentiable Image Formation Model with a HARD Straight Thought Estimator (STE) selection mask, unrolled Iterative Soft-Thresholding Algorithm (ISTA) deconvolution, and a residual Convolutional Neural Network (CNN).  
The objective combines physical consistency (Point Spread Function (PSF) and convolutional formation model) with structural fidelity (contrast, Side-Lobe-Ratio (SLR), entropy, and row diversity).  
Simulations using a 3.5\,MHz probe show that the learned configuration preserves axial and lateral resolution with half of the active elements.  
This physics-guided, data-driven approach enables compact, cost-efficient ultrasound probe design without sacrificing image quality, and it is expandable to 3-D volumetric imaging.
\end{abstract}
%
\section{Introduction}
\label{sec:intro}
Ultrasound imaging offers real-time, non-invasive visualization for medical and industrial applications, but dense transducer arrays remain costly and power-demanding.  
A conventional linear probe consists of dozens to hundreds of piezoelectric elements operating in both transmission and reception, with typical pitches near half a wavelength to avoid grating lobes. 
Modern 2-D arrays for volumetric imaging may include thousands of elements, leading to high hardware and bandwidth demands~\cite{ramalli2022design}.  
Reducing the number of active elements while preserving spatial resolution and contrast therefore remains a key challenge in probe design.
Sparse arrays address this issue by reducing the number of active elements or activating only a subset of a full aperture.  
Existing strategies rely on deterministic layouts, such as spiral, aperiodic, or Vernier patterns~\cite{martinez20102d}, or stochastic schemes based on random sampling and metaheuristics~\cite{trucco2002thinning}.  
While these methods lower hardware complexity, they decouple probe geometry optimization from image formation and require iterative post-design evaluation.  
Recent studies have linked array configuration and imaging physics through inverse-problem formulations \cite{goudarzi2023unifying}, while differentiable modeling now enables joint optimization of hardware and reconstruction \cite{simson2023differentiable,arguello2023deep}.  
Deep networks have also been applied to estimate probe geometry~\cite{noda2022flexible,qi2023deep} or compensate for target-dependent variability~\cite{zhang2022targetaware}, yet these typically treat reconstruction and geometry separately.
In contrast to the existing literature, this paper learns the array configuration itself within a differentiable forward model, enabling physics-guided probe design rather than post-acquisition calibration.  
The proposed framework integrates a differentiable simulator with a learnable sparse-array selection based on the Hard Straight-Through Estimator (STE)~\cite{yin2019understanding}, coupled with unrolled ISTA deconvolution and a residual CNN head~\cite{hyun2021deep}.  
By coupling physics-based beamforming (PyMUST/SIMUS)~\cite{garcia2022simus} with deep reconstruction, gradients propagate from image-level losses back to transducer activation, allowing end-to-end learning of optimal sparse configurations.  
\textcolor{black}{Although demonstrated with a 1-D linear probe in a controlled simulation setting, the same differentiable design principle extends directly to 2-D matrix arrays, paving the way for adaptive 3-D volumetric imaging; broader validation on other probes and tissue-mimicking phantoms is left for future work.}

\vspace{-4mm}
\section{Image Formation Model}
\label{sec:Forward}
Ultrasound image formation can be approximated by a linear convolution between the reflectivity (scatterer) map and the system point-spread function (PSF).  
This linear model assumes weak scattering and small phase variations, i.e., the Born approximation holds. \textcolor{black}{A shift-invariant PSF is assumed to keep the end-to-end optimization tractable.}
This work model the acquisition at a given imaging depth as:
\begin{equation}
\mathbf{Y_c}= \mathcal{F}^{-1}\!\big(\mathcal{F}(\bm{\kappa_c})\odot\mathcal{F}(\mathbf{I_s})\big),
\label{eq:forw}
\end{equation}
where $\mathbf{I_s}$ is the scatterer representation of the speckle-free reference image $\mathbf{I_{ref}}$, defined as a 2D matrix encoding the spatial distribution of point scatterers over a grid with specified size and resolution.  
Each nonzero entry in $\mathbf{I_s}$ corresponds to a point scatterer at the associated grid position, whose amplitude represents the local acoustic reflectivity.  
$\bm{\kappa_c}$ denotes the complex-valued PSF kernel, $\mathcal{F}$ the 2D Fourier transform, and $\odot$ the elementwise (Hadamard) product.  
Throughout the paper, the subscript $c$ denotes complex-valued variables.
\section{Sparse Mask Learning via HARD-STE}
\label{sec:sparsemask}
Consider a linear probe with $N_e$ available elements, from which we aim to select $k$ active ones ($k < N_e$). In this work, the same elements are considered for both emission and reception modes.  
The sparse transmit/receive selection is modeled by a binary matrix 
$\mathbf{P}\!\in\!\{0,1\}^{N_e\times k}$, where each column activates one element of the probe.
\textcolor{black}{Selection is performed sequentially without replacement: each column is one-hot and previously selected elements are masked out, yielding exactly $k$ distinct active elements in $\mathbf{P}_{\text{hard}}$.}
The selected coordinates are computed as $\mathbf{x}_e^{(s)} = \mathbf{x}_e \mathbf{P}$, $\mathbf{z}_e^{(s)} = \mathbf{z}_e \mathbf{P}$, and $\mathbf{h}^{(s)} = \mathbf{h} \mathbf{P}$, where $\mathbf{x}_e$, $\mathbf{z}_e$, and $\mathbf{h}$ are row vectors of length $N_e$ representing the lateral ($x$) and axial ($z$) positions of the probe elements and apodization weights of the full probe with all active elements. Multiplying by $\mathbf{P}$ selects the subset of $k$ active elements used to form the sparse aperture.
The selected parameters are passed to  SIMUS~\cite{garcia2022simus}, a state-of-the-art ultrasound simulator, to generate the corresponding PSF $\bm{\kappa_{c}}$, which implicitly depends on the current selection mask $\mathbf{P}$.  
This formulation embeds the element-selection process directly into the differentiable simulator, enabling physically consistent aperture learning.  
Because direct optimization of a binary mask is non-differentiable, the matrix $\mathbf{P}$ is parameterized by continuous logits $\bm{\ell}\!\in\!\mathbb{R}^{N_e\times k}$, from which columnwise probabilities are obtained using a temperature-controlled softmax:
$\mathbf{p}_j=\mathrm{softmax}(\bm{\ell}_j/\tau)$ with $\tau>0$.  
During training, high temperatures produce smooth probability maps that allow gradient flow, while progressively decreasing $\tau$ sharpens them toward one-hot selections.  
To combine discrete inference with differentiable learning, a Hard Straight-Through Estimator (HARD-STE) is applied:
\begin{equation}
\mathbf{P} = (\mathbf{P}_{\text{hard}} - \mathbf{P}_{\text{soft}})_{\text{detach}} + \mathbf{P}_{\text{soft}},
\end{equation}
where $\mathbf{P}_{\text{soft}}=[\mathbf{p}_1,\dots,\mathbf{p}_k]$ carries gradients and $\mathbf{P}_{\text{hard}}$ represents the corresponding binary selections.  
The \texttt{detach} operator disconnects the non-differentiable hard assignment from the computation graph, preventing gradient flow through $\mathbf{P}_{\text{hard}}$ while allowing updates to propagate through $\mathbf{P}_{\text{soft}}$.  
\textcolor{black}{This mechanism yields discrete behavior in the forward pass and continuous gradients during backpropagation, while $\mathcal{L}_{\text{row}}$ tries to avoid duplicate selections across columns.}
\section{End-To-End Optimization}
\label{sec:typestyle}
\textcolor{black}{The framework jointly optimizes the selection mask and reconstruction parameters within a differentiable pipeline, so that image-level losses backpropagate through the simulator to the mask:
}
\begin{equation}
    \hat{\textbf{P}},\hat{\boldsymbol{\theta}}=\min_{\mathbf{P},\,\boldsymbol{\theta}} \; \mathcal{L}\big(f_{\boldsymbol{\theta}}(Y_{\mathbf{c}}),\,\mathbf{I_{\mathrm{ref}}}\big),
\end{equation}
where $\hat{\mathbf{P}}$ is the optimal selection mask and $\hat{\boldsymbol{\theta}}$ the optimal weights of the recovery algorithm. The recovery algorithm is composed of the unrolled ISTA and a residual CNN head. 

The PSF kernels of the sparse probe $\bm{\kappa_{\mathbf{c}}}$ and the full probe $\bm{\kappa_{\text{ref}}}$ are synthesized using the differentiable SIMUS~\cite{garcia2022simus} simulator configured for plane-wave imaging with seven angles uniformly sampled from $-10^{\circ}$ to $10^{\circ}$.  
A single centered scatterer is placed on the imaging grid, and the received pressure field is computed with the Pfield routine based on the Rayleigh-Sommerfeld integral and a linear spatial impulse response, using as input the transmitter and receiver positions, apodization, and transmit delays.  
The resulting RF signals are quadrature-demodulated into complex baseband (IQ) form, beamformed via a delay-and-sum operator, and windowed to $128{\times}128$ before normalization to produce the PSF kernel $\bm{\kappa_c}$ used in training.
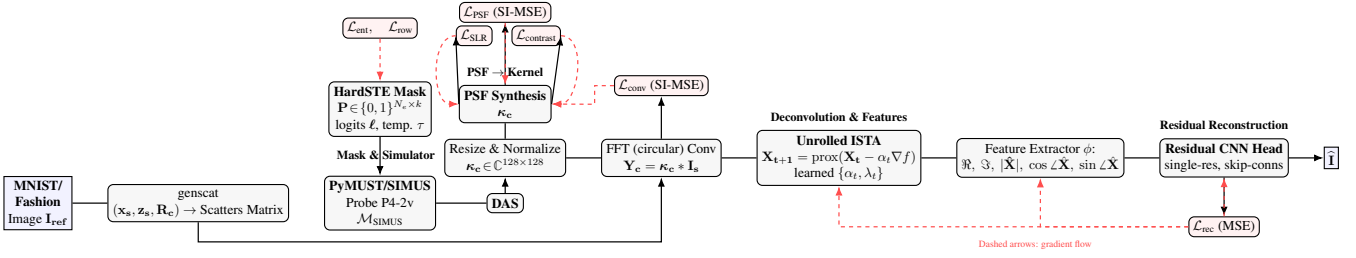
\begin{figure*}[ht]
\centering

\resizebox{1\linewidth}{!}{%
\begin{tikzpicture}[
  node distance=4mm and 9mm,
  >=Latex,
  proc/.style={draw, rounded corners, thick, align=center, inner sep=3.5pt, fill=gray!6},
  loss/.style={draw, rounded corners, thick, align=center, inner sep=3pt, fill=red!5},
  data/.style={draw, thick, align=center, inner sep=3pt, fill=blue!5},
  arr/.style={->,line width=0.9pt},
  back/.style={->,line width=0.9pt, dashed, red!70},
  tit/.style={font=\small\bfseries}
]

\node[data] (img) {\textbf{MNIST/}\\\textbf{Fashion}\\Image $\mathbf{I_{ref}}$};
\node[proc, right=of img] (genscat) {genscat \\ $(\mathbf{x_s},\mathbf{z_s}, \mathbf{R_c})$ $\rightarrow$ Scatters Matrix};
\node[proc, right=of genscat] (simus) {\textbf{PyMUST/SIMUS}\\Probe P4-2v\\$\mathcal{M}_\text{SIMUS}$};

\node[proc, above=of simus, yshift=6mm] (mask) {\textbf{HardSTE Mask}\\$\mathbf{P}\!\in\!\{0,1\}^{N_e\times k}$\\logits $\bm{\ell}$, temp. $\tau$};

\node[proc, right=of simus, xshift=4mm] (das) {\textbf{DAS}};
\node[proc, above=of das, yshift=14mm] (psf) {\textbf{PSF Synthesis}\\ $\mathbf{\bm\kappa_\mathbf{c}}$};
\node[proc, below=of psf] (kern) {Resize \& Normalize\\$\bm{\kappa_{\mathbf{c}}}\!\in\!\mathbb{C}^{128\times128}$};

\node[proc, right=of kern] (conv) {FFT (circular) Conv\\$\mathbf{Y_c} = \mathbf{\bm\kappa_{\mathbf{c}}} * \mathbf{I_{s}}$};

\node[proc, right=of conv] (ista) {\textbf{Unrolled ISTA}\\$\mathbf{X_{t+1}}=\mathrm{prox}(\mathbf{X_t}-\alpha_t\nabla f)$\\learned $\{\alpha_t,\lambda_t\}$};
\node[proc, right=of ista] (phi) {Feature Extractor $\phi$:\\$\Re,\ \Im,\ |\mathbf{\hat{X}}|,\ \cos\angle\hat{\mathbf{X}},\ \sin\angle\hat{\mathbf{X}}$};

\node[proc, right=of phi] (cnn) {\textbf{Residual CNN Head}\\single-res, skip-conns};
\node[data, right=of cnn] (yhat) {$\widehat{\mathbf{I}}$};

\node[loss, above=of psf, yshift=12mm] (lpsf) {$\mathcal{L}_{\text{PSF}}$ (SI-MSE)};
\node[loss, above=of conv, yshift=7mm] (lconv) {$\mathcal{L}_{\text{conv}}$ (SI-MSE)};
\node[loss, below=of cnn, yshift=-6mm] (lrec) {$\mathcal{L}_{\text{rec}}$ (MSE)};
\node[loss, above=of psf, yshift=6mm, xshift=8mm] (lcontr) {$\mathcal{L}_{\text{contrast}}$};
\node[loss, above=of psf, yshift=6mm, xshift=-8mm] (lslr) {$\mathcal{L}_{\text{SLR}}$};
\node[loss, above=of mask, yshift=7mm] (lent) {$\mathcal{L}_{\text{ent}}, \quad \mathcal{L}_{\text{row}}$};

\draw[arr] (img) -- (genscat) -- ++(0,-1) -| (conv);
\draw[arr] (mask) -- (simus);
\draw[arr] (simus) -- (das) -- (kern);
\draw[arr] (psf) -- (kern) -- (conv) -- (ista) -- (phi) -- (cnn) -- (yhat);

\draw[arr] (psf) -- (lpsf);
\draw[arr] (conv) -- (lconv);
\draw[arr] (cnn) -- (lrec);
\draw[arr] (psf.east) -- (lcontr.east);
\draw[arr] (psf.west) -- (lslr.west);
\node[font=\scriptsize, red!70, below=15mm of phi, xshift=-1.3mm] {Dashed arrows: gradient flow};
\draw[back] (lrec.north) -- (cnn.south);
\draw[back] (lrec.west)  -| (phi.south) ;
\draw[back] (lrec.west)  -| (ista.south);
\draw[back] (lconv.west) -- ++(-0.5,0) -- ++(0,-0.48) -- (psf.east);
\draw[back] (lpsf.south)  to[out=-90,in=90] (psf.north);
\draw[back] (lcontr.east) to[out=0,in=0] (psf.east);
\draw[back] (lslr.west)  to[out=180,in=180] (psf.west);
\draw[back] (lent.south)  -| (mask.north);

\node[tit, above=3mm of simus, xshift=1.5mm] (cap1) {Mask \&    Simulator};
\node[tit, above=1mm of psf]   (cap2) {PSF $\rightarrow$ Kernel};
\node[tit, above=1mm of ista]  (cap3) {Deconvolution \& Features};
\node[tit, above=1mm of cnn]   (cap4) {Residual Reconstruction};

\end{tikzpicture}

}
\vspace{-8mm}
\caption{Overview of the proposed end-to-end ultrasound optimization framework.
In the forward path, the simulator generates synthetic RF data based on the active elements selected by $\mathbf{P}$, which are then reconstructed by the unrolled ISTA and CNN modules to match a reference image. Gradients from the reconstruction loss propagate backward through the network and simulator to update both the reconstruction parameters and the selection mask, enabling physics-guided sparse-array learning directly from image-level supervision.}
\label{fig:Main}
\vspace{-5mm}
\end{figure*}
A schematic overview is shown in Fig.\ref{fig:Main}.
\subsection{Unrolling ISTA Algorithm for Deconvolution}
The unrolled ISTA module acts as a physics-inspired deconvolution stage that iteratively removes PSF-induced blurring from the complex RF image $\mathbf{Y_c}$ while enforcing sparsity in the reconstructed reflectivity map.  
It emulates an iterative inverse-problem solver whose step sizes $\{\alpha_t\}$ and thresholds $\{\lambda_t\}$ are learned through backpropagation for faster, data-adaptive convergence.  
Formally, given $\mathbf{Y_c}$ and the PSF kernel $\bm{\kappa_c}$, the unrolled ISTA performs $n_{\text{ISTA}}$ proximal-gradient iterations:
\begin{equation}
\mathbf{X_{t+1}} = \mathrm{prox}_{\lambda_t}\!\left(\mathbf{X_t} - \alpha_t\,\mathbf{A}^\dagger(\mathbf{A}\mathbf{X_t} - \mathbf{Y_c})\right),
\end{equation}
where $\mathbf{A}$ is the convolution operator defined by Eq.~\eqref{eq:forw}.  
The proximal operator applies complex soft-thresholding,
\begin{equation}
\textcolor{black}{\mathrm{prox}_{\lambda}(\mathbf{Z})
= \frac{\mathbf{Z}}{|\mathbf{Z}|+\varepsilon}\max(|\mathbf{Z}|-\lambda,0), \quad \varepsilon>0.
}
\end{equation}
and its parameters are jointly optimized with the rest of the network.

\subsection{Reconstruction Head residual CNN}
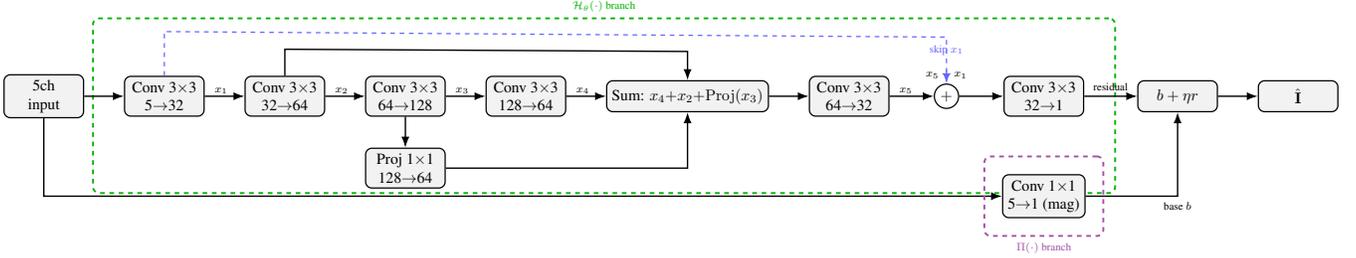
\begin{figure*}[ht]
\centering
\resizebox{\linewidth}{!}{%
\begin{tikzpicture}[
  >=Latex, node distance=7mm and 9mm,
  blk/.style={draw, thick, rounded corners, minimum width=18mm, minimum height=7mm, align=center, fill=gray!10},
  conn/.style={->, line width=0.9pt},
  skip/.style={-Latex, dashed, line width=0.8pt, blue!60},
  lab/.style={font=\scriptsize},
  op/.style={circle, draw, thick, minimum size=5.5mm, inner sep=0pt, fill=white}
]
\node[blk] (in) {5ch\\input};
\node[blk, right=of in] (e1a) {Conv 3$\times$3\\5$\rightarrow$32};
\node[blk, right=of e1a] (e1b) {Conv 3$\times$3\\32$\rightarrow$64};
\node[blk, right=of e1b] (e2) {Conv 3$\times$3 \\64$\rightarrow$128};
\node[blk, right=of e2] (e3) {Conv 3$\times$3\\128$\rightarrow$64};

\node[blk, below=7mm of e2] (proj) {Proj 1$\times$1\\128$\rightarrow$64};
\node[blk, right=of e3] (mix) {Sum: $x_4{+}x_2{+}\mathrm{Proj}(x_3)$};
\node[blk, right=of mix] (c5) {Conv 3$\times$3\\64$\rightarrow$32};

\node[lab, anchor=west] at ($(e1a.east)+(1mm,1.3mm)$) {$x_1$};
\node[lab, anchor=west] at ($(e1b.east)+(1mm,1.3mm)$) {$x_2$};
\node[lab, anchor=west] at ($(e2.east)+(1mm,1.3mm)$)  {$x_3$};
\node[lab, anchor=west] at ($(e3.east)+(1mm,1.3mm)$)  {$x_4$};
\node[lab, anchor=west] at ($(c5.east)+(1mm,1.3mm)$)  {$x_5$};

\node[op, right=10mm of c5] (plus) {$+$};
\node[lab, above=0mm of plus] {$x_5 + x_1$};

\node[blk, right=10mm of plus] (head) {Conv 3$\times$3\\32$\rightarrow$1};
\node[blk, below=of head, yshift=-6mm] (base) {Conv 1$\times$1\\5$\rightarrow$1 (mag)};
\node[blk, right=12mm of head] (sum) {$b + \eta r$};
\node[blk, right=of sum] (out) {$\hat{\mathbf{I}}$};

\draw[conn] (in) -- (e1a);
\draw[conn] (e1a) -- (e1b);
\draw[conn] (e1b) -- (e2);
\draw[conn] (e2) -- (e3);

\draw[conn] (e2) -- (proj);
\draw[conn] (e3) -- (mix);

\coordinate (top1) at ($(e1b.north)+(0,6mm)$);
\coordinate (top2) at ($(mix.north)+(0,6.5mm)$);

\draw[conn] (e1b) -- (top1) -- (top2) -- (mix);
\draw[conn] (proj) -| (mix);

\draw[conn] (mix) -- (c5);
\draw[conn] (c5) -- (plus);

\coordinate (topA) at ($(e1a.north)+(0,10mm)$);
\coordinate (topB) at ($(plus.north)+(0,10.5mm)$);
\draw[skip] (e1a.north) -- (topA) -- (topB) -- node[lab,above]{skip $x_1$} (plus.north);

\draw[conn] (plus) -- (head);
\draw[conn] (head) -- node[lab,above]{residual} (sum);

\draw[conn] (in) |- (base);
\draw[conn] (base) -| node[lab,below]{base $b$} (sum);
\draw[conn] (sum) -- (out);


\node[draw=violet!70, very thick, dashed, rounded corners,
      fit=(base), inner sep=4mm, label={[violet!70]below:{\scriptsize $\Pi(\cdot)$ branch}}] (PiBox) {};

\begin{scope}[on background layer]
\node[draw=green!70!black, very thick, dashed, rounded corners,
      fit=(e1a)(e1b)(e2)(e3)(proj)(mix)(c5)(plus)(head),
      inner sep=7mm, 
       yshift=6mm, 
      label={[green!70!black]above:{\scriptsize $\mathcal{H}_{\theta}(\cdot)$ branch}}] (HthetaBox) {};
      \end{scope}
\end{tikzpicture}
}

\vspace{-4mm}
\caption{Residual CNN reconstruction head combining a projection branch $b=\Pi(|\hat{\mathbf{x}}|)$ and a residual refinement branch $\mathcal{H}_\theta(\cdot)$. The fused feature $x_5+x_1$ is mapped to a residual image $r$ by the final convolutional block, and the final output is $\widehat{\mathbf{I}} = b + \eta r$.}
\label{fig:CNN}
\vspace{-4mm}
\end{figure*}

After unrolled ISTA, the estimated complex field $\hat{\mathbf{X}}$ is converted into five real-valued features:

\noindent $\phi(\hat{\mathbf{X}}) = [\Re(\hat{\mathbf{X}}),\,\Im(\hat{\mathbf{X}}),\,|\hat{\mathbf{X}}|,\,\cos(\angle\hat{\mathbf{X}}),\,\sin(\angle\hat{\mathbf{X}})],$ which feeds a single-scale residual CNN head.
The reconstruction mapping can be summarized as:

\[
\widehat{\mathbf{I}} = \Pi(\mathbf{\phi(\hat{\mathbf{X}})}) + \eta\,\mathcal{H}_\theta(\mathbf{\phi(\hat{\mathbf{X}})}),
\]
where $\Pi(\cdot)$ is a $1{\times}1$ base projection from the magnitude channel, and 
$\mathcal{H}_\theta(\cdot)$ is the residual CNN with three skip connections.
The network uses $B {=} 32$ initial filters, and each convolutional layer applies a LeakyReLU activation with slope $\sigma {=} 0.1$. The architecture is shown in Fig.\ref{fig:CNN}.

The final reconstruction $\widehat{\mathbf{I}}$ combines the output of the unrolled ISTA and the residual CNN head, producing a refined estimate that recovers spatial details from the blurred input $Y_c$ using the learned PSF kernel $\bm{\kappa_c}$.
\subsection{Loss Function}
\label{sec:loss}
The training objective combines physical consistency with image-level fidelity.  
The total loss is expressed as:
\begin{equation}
\begin{split}
 &   \mathcal{L}_{\text{total}}
= w_{\text{PSF}}\mathcal{L}_{\text{PSF}}
+ w_{\text{conv}}\mathcal{L}_{\text{conv}}
+ w_{\text{rec}}\mathcal{L}_{\text{rec}} \\
&  \quad + \lambda_{\text{contrast}}\mathcal{L}_{\text{contrast}}
+ \lambda_{\text{SLR}}\mathcal{L}_{\text{SLR}}
+ \lambda_{\text{ent}}\mathcal{L}_{\text{ent}}
+ \lambda_{\text{row}}\mathcal{L}_{\text{row}}. \nonumber
\end{split} 
\end{equation}
The total objective combines physical consistency, image fidelity, and PSF regularization terms.

\noindent The PSF loss $\mathcal{L}_{\text{PSF}}=\mathrm{SI\text{-}MSE}(\bm\kappa_{\mathbf{c}},\bm\kappa_{\text{ref}})$ with $\mathrm{SI\text{-}MSE}(\hat{\mathbf{Y}},\mathbf{Y})=\|\alpha^*\hat{\mathbf{Y}}-\mathbf{Y}\|_2^2/\|\mathbf{Y}\|_2^2$ and $\alpha^*=\langle\hat{\mathbf{Y}},\mathbf{Y}\rangle/\langle\hat{\mathbf{Y}},\hat{\mathbf{Y}}\rangle$ enforces physical similarity between the synthesized from the learned mask and reference PSFs, independent of the input data.  
In contrast, the convolution loss $\mathcal{L}_{\text{conv}}=\|\mathbf{Y}_{\mathbf{c}}-\mathbf{Y}_{\text{ref}}\|_2^2,$ evaluates the consistency of the blurred observations produced by the forward model, allowing gradients to reach the selection mask through fewer intermediate layers, where $\mathbf{Y_c}$ and $\mathbf{Y}_{\text{ref}}$ denote the blurred complex fields obtained by convolving the scatterer map with $\bm\kappa_{\mathbf{c}}$ and $\bm\kappa_{\text{ref}}$, respectively.  
Although both terms are correlated—since similar PSFs imply similar convolutions—their combination stabilizes training by coupling physics-level supervision with data-dependent feedback.

\noindent The reconstruction loss $\mathcal{L}_{\text{rec}}=\|\widehat{\mathbf{I}}-\mathbf{I}_{\mathrm{ref}}\|_2^2,$ penalizes image-level deviations, where $\mathbf{I_{\mathrm{ref}}}$ is the reference image used to generate the scatterer map $\mathbf{I_s}$ in Eq.~\eqref{eq:forw}. 

\noindent PSF Regularization terms are computed on the normalized PSF magnitude $|\bm\kappa_c|$:  
the \emph{contrast loss} $\mathcal{L}_{\text{contrast}}=(\tfrac{1}{|\Omega_{\text{out}}|}\sum_{(x,z)\in\Omega_{\text{out}}}|\bm\kappa_c|^2)/(\tfrac{1}{|\Omega_{\text{in}}|}\sum_{(x,z)\in\Omega_{\text{in}}}|\bm\kappa_c|^2)$ reduces background energy outside the focal region.

\noindent Two complementary sidelobe ratio losses are defined as $\mathcal{L}_{\text{SLR-q}}=A_{\text{side}}/A_{\text{main}}$, with $A_{\text{main}}=\max_{r\le r_{\text{main}}}|\bm\kappa_c|$ and
$A_{\text{side}}=\mathrm{quantile}_q(|\bm\kappa_c(r)|,\,r\ge r_{\text{guard}})$,  
and $\mathcal{L}_{\text{SLR-i}}=\mathrm{E}_{\text{side}}/\mathrm{E}_{\text{main}}$, where $\mathrm{E}_{\text{main}}=\sum_{r\le r_{\text{main}}}|\bm\kappa_c|^2$, and $\mathrm{E}_{\text{side}}=\sum_{r\ge r_{\text{guard}}}|\bm\kappa_c|^2$. $\mathcal{L}_{\text{SLR}} = \tfrac{1}{2}\!\left(\mathcal{L}_{\text{SLR-q}}+\mathcal{L}_{\text{SLR-i}}\right),$
where $r_{\text{main}}\!=\!1.5\lambda_0$ and $r_{\text{guard}}\!=\!2.5\lambda_0$.
Together, $\mathcal{L}_{\text{contrast}}$ and $\mathcal{L}_{\text{SLR}}$ promote compact PSFs with high contrast and low sidelobes.
\textcolor{black}{$\mathcal{L}_{\text{ent}}$ penalizes diffuse softmax columns to promote confident selections, while $\mathcal{L}_{\text{row}}$ penalizes repeated element usage across columns to encourage $k$ distinct active elements; the weights are empirically tuned for stable training.
}
\vspace{-2mm}
\section{Experimental Setup}
\label{sec:majhead}
The proposed framework was trained using MNIST and FashionMNIST as input domains.  These datasets serve as controlled scatterer distributions, offering reproducible test fields with varying spatial complexity while avoiding the variability of tissue-dependent backscattering.
Each image is converted into a scatterer distribution using \texttt{genscat} function from Pymust, with scatterers positioned at depths $z\!\in\![0,15]$\,cm.  
The forward model is based on the PyMUST \texttt{getparam('P4-2v')} probe, comprising $N_e\!=\!64$ elements, of which $k\!=\!32$ are active.  
Acoustic properties are defined by a center frequency $f_c\!=\!\SI{3.5}{MHz}$, sound speed $c\!=\!\SI{1540}{m/s}$, and sampling frequency $f_s\!=\!4f_c\!=\!\SI{14}{MHz}$, yielding a wavelength $\lambda_0\!=\!c/f_c\!=\!\SI{0.44}{mm}$ and grid pitch $p\!=\!(c/f_s)/4\!=\!\SI{27.5}{\micro\meter}$.  
The imaging field spans $x\!\in\![-p n_x/2,\,p n_x/2]$ and $z\!\in\![0,\,p n_z]$ with $(n_x,n_z)\!=\!(400,400)$, from which the complex PSF $\bm\kappa_c\!\in\!\mathbb{C}^{128\times128}$ is cropped and energy-normalized for FFT-based convolutions.  

Training is performed for $100$ epochs with batch size $64$ using the Adam optimizer, with learning rates $\eta_{\text{mask}}\!=\!10^{-4}$, $\eta_{\text{ISTA}}\!=\!5{\times}10^{-4}$, and $\eta_{\text{head}}\!=\!10^{-4}$.  
Loss weights are $w_{\text{PSF}}\!=\!10$, $w_{\text{conv}}\!=\!1$, $w_{\text{rec}}\!=\!10$, $\lambda_{\text{contrast}}\!=\!\lambda_{\text{SLR}}\!=\!10$, $\lambda_{\text{ent}}\!=\!1$, and $\lambda_{\text{row}}\!=\!10$.  
The softmax temperature $\tau$ follows a cosine decay from $\tau_0\!=\!6$ to $1.2$ over $T_{\text{warm}}\!=\!1500$ steps.  
ISTA unrolling uses $n_{\text{ISTA}}\!=\!8$ iterations with trainable $\{\alpha_t,\lambda_t\}$ per layer.
\section{Results}
\label{sec:results}
\begin{figure}[ht]
\centering
\includegraphics[width=\linewidth]{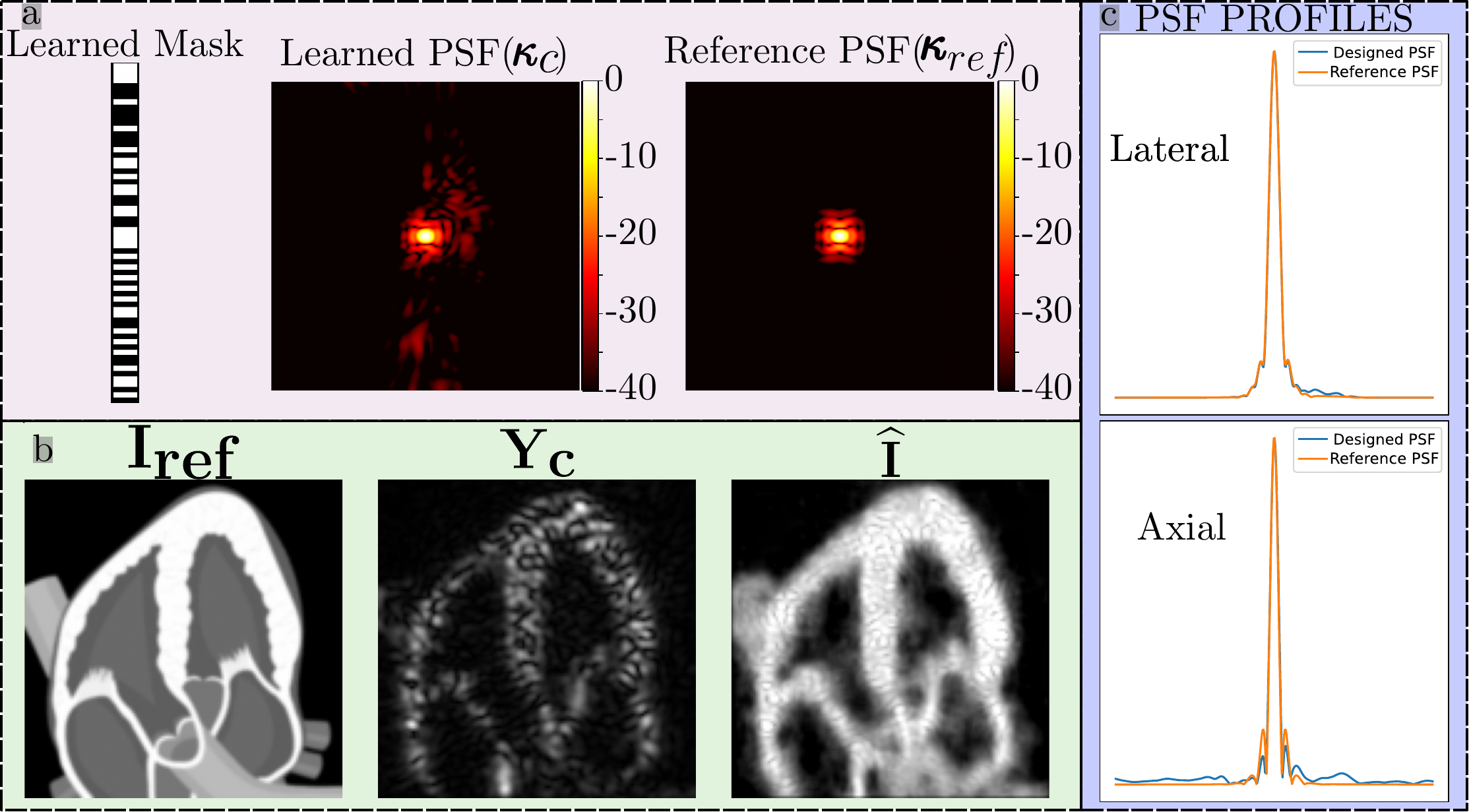}
\caption{(Comparison between the learned sparse configuration and the full-aperture reference.
(\textbf{a}) Learned binary mask from the matrix $\mathbf{\hat{P}}$ showing the active transmit/receive elements (left), its corresponding synthesized PSF $\bm\kappa_{\mathbf{c}}$ from the differentiable simulator using the optimized mask (center), and the full-aperture reference PSF $\bm\kappa_{\text{ref}}$ (right). All PSFs are displayed on a \SI{40}{dB} logarithmic dynamic range.
(\textbf{b}) Ground-truth image $\mathbf{I_{ref}}$ (left), blurred measurement $\mathbf{Y_c}$ (middle), and reconstructed image $\widehat{\mathbf{I}}$ obtained after unrolled ISTA and CNN refinement (right). The learned configuration achieves comparable axial and lateral resolution while using only half of the active elements.
(\textbf{c}) Lateral (top) and axial (bottom) profiles comparing the learned and reference PSFs, showing close main-lobe widths and reduced sidelobe levels for the learned design.
}
\label{fig:results}
\vspace{-1mm}
\end{figure}

The proposed framework enables gradient flow from the image reconstruction loss back to the transmit/receive selection pattern.  
Fig.~\ref{fig:results} shows the optimized mask, reference PSF, learned PSF, and reconstructed image.  
\textcolor{black}{Despite using only half of the active elements, the learned configuration preserves axial and lateral resolution and yields $k$ distinct active elements in the reported runs.}
For the reconstructed image, to assess a clinical-type image, a fine-tuning of the Reconstruction Head residual CNN was performed for 5 epochs.
Quantitative results in Table~\ref{tab:quant} compare uniform, by removing every odd element, a random not learned distribution, choosing the best distribution out of random 500.000 tries, and learned configurations against the full array.  

The proposed method achieves PSF quality and contrast close to the dense configuration while using 50\% fewer elements.  
Deterministic and stochastic baselines were also evaluated, including a uniform mask activating every second element and a random mask chosen once without learning.  
\textcolor{black}{
These fixed configurations provide interpretable reference designs, showing that end-to-end optimization yields improved PSF compactness and contrast over non-optimized layouts, while preserving axial and lateral resolution and yielding $k$ distinct active elements in the reported runs.}
The composite metric $\bar{\mathcal{L}} = \tfrac{1}{4}(\mathcal{L}_{\text{PSF}} + \mathcal{L}_{\text{contrast}} + \mathcal{L}_{\text{SLR-q}} + \mathcal{L}_{\text{SLR-i}})$ confirms that the learned configuration approaches full-array performance in both mean error and sidelobe suppression.

\vspace{-4mm}
\begin{table}[ht]
\centering
\caption{Quantitative comparison of probe configurations. Lower is better.}
\resizebox{\linewidth}{!}{
\begin{tabular}{lccccc}
\toprule
\textbf{Config.} &
$\mathcal{L}_{\text{PSF}}$ &
$\mathcal{L}_{\text{contrast}}$ &
$\mathcal{L}_{\text{SLR-q}}$ &
$\mathcal{L}_{\text{SLR-i}}$ &
Mean \\
\midrule
Uniform & 0.0522 & 3.743e--4 & 0.0249 & 0.2020 & 0.0699 \\
Random  & 0.0736 & 3.258e--4 & 0.0196 & 0.1528 & 0.0616 \\
Full    & \underline{0.0000} & \underline{1.766e--4} & \underline{0.0069} & \underline{0.1349} & \underline{0.0355} \\
\textbf{Proposed} & \textbf{0.0661} & \textbf{1.900e--4} & \textbf{0.0179} & \textbf{0.1452} & \textbf{0.0573} \\
\bottomrule
\end{tabular}
}
\label{tab:quant}
\end{table}

\vspace{-6mm}
\section{Conclusion}
\label{sect:conclusion}
\vspace{-2mm}
This work introduced an end-to-end framework that jointly optimizes sparse ultrasound probe design and image reconstruction.  
By coupling a differentiable forward model with an unrolled ISTA and residual CNN, the method enables gradient-based learning of transmit/receive configurations that preserve PSF quality and contrast using only half of the active elements.  
While demonstrated on a 1-D linear probe, the same differentiable design principle naturally extends to 2-D matrix arrays, enabling adaptive 3-D volumetric imaging through joint hardware-software optimization.  
These results highlight the potential of physics-informed differentiable modeling for the design of compact, high-quality ultrasound systems.
\textcolor{black}{The approach increases offline training cost but keeps inference practical by using a fixed learned mask and a lightweight reconstruction network, shifting complexity from hardware to software.}
\vspace{-4mm}
\section{Compliance with ethical standards}
\label{sec:ethics}
This is a numerical simulation study for which no ethical approval was required. 
\section{Acknowledgments}
\label{sec:acknowledgments}
This work was supported by the STIC-AmSud Project CTO MINCIENCIAS 110-2024: Computational ultrasound imaging: from theory to applications-COMPOUND.
This work was supported by the LABEX CELYA (ANR-10-LABX-0060) and LABEX PRIMES (ANR-11-LABX-0063) of Université de Lyon, within the program ”Investissements d’Avenir” (ANR-11-IDEX-0007) operated by the French National Research Agency (ANR), as well as the ”CAVI-IAR” Project (ANR-22-CE19-0006), operated by the French National
Research Agency (ANR).
\bibliographystyle{IEEEbib}
\bibliography{refs}
\end{document}